\documentclass[12pt,a4paper]{article}
\usepackage{amsfonts}
\usepackage{amsmath}
\usepackage[ top=2cm , bottom=2cm , left=2cm , right=2cm ]{geometry}

\input{tcilatex}
\begin{document}

\begin{titlepage}
\vspace*{3cm}
\begin{center}
{\Large \textsf{\textbf{New treatment of red/blue shifts of emitted photons in terms of Kerr-Newman black hole parameters}}}
\end{center}
\vskip 5mm
\begin{center}
{\large \textsf{Houssam Eddine Trad$^{a}$ and Lamine Khodja$^{a}$\footnote{Email adress: lamine.khodja@yahoo.fr, khodja.lamine@univ-ouargla.dz}}}\\
\vskip 5mm
$^{a}$Laboratoire de Rayonnement et Plasmas et Physique de Surfaces \\
D\'{e}partement de Physique, Facult\'{e} des Math\'{e}matiques et des
Sciences de la Mati\'{e}re, Universit\'{e} Kasdi Merbah - Ouargla, 30000 Ouargla, Algeria.\\
\end{center}
\vskip 2mm
\begin{center}{\large\textsf{\textbf{Abstract}}}\end{center}
\begin{quote}
This study presents a method for determining the mass, angular momentum, and charge of Kerr-Newman black holes by analyzing the red/blue shifts of photons emitted by geodesic neutral massive objects. We derive the equations of motion for photons in the Kerr-Newman spacetime. We  obtain explicit expressions for redshift/blueshift photons emitted by an object orbiting a Kerr-Newman black  hole. Finally,  using the Boyer-Lindquist coordinates, we express the red/blue shifts in terms of the metric parameters.
\end{quote}
\vspace*{2cm}
\noindent\textbf{\sc Pacs numbers}: 97.60.Lf, 04.70.-s, 98.62.Py
\end{titlepage}

\section{Introduction}

In 4-dimensional General Relativity, the no hair theorem states that all
black hole solutions to the Einstein-Maxwell equations are uniquely
characterized by three numbers: mass $M$, electric charge $Q$, and angular
momentum $J$. The simplest case is the neutral static Schwarzschild
solution.The Reissner-Nordstr\"{o}m solution is a generalization of the
Schwarzschild solution to the charged case, while the Kerr solution is the
generalization to the spinning case \cite{Kerr}. The generalization of the
Schwarzschild solution to the charged and spinning case is the Kerr-Newman
metric \cite{Newman}. Although the charge aspect of black hole has not
important astrophysical relevance \cite{Adamo}, there are scenarios where
the collapse of compact stars can lead to the formation of a Reissner-Nordstr%
\"{o}m \cite{Ray} or Kerr-Newman black holes \cite{Nathanail}. Moreover, the
fact that the Kerr-Newman metric is the most general stationary black hole
solution to the Einstein-Maxwell equations gives it a great importance for
theoretical considerations and the understanding of other black holes \cite%
{Adamo}\cite{Kraniotis}.

Astrophysical observations are suggesting the existence of a supermassive
black hole at the center of many spiral galaxies, such as Sagittarius A*
(SgrA*), the supermassive black hole at the center of the Milky Way galaxy 
\cite{Begelman}\cite{Shen}\cite{Ghez}\cite{Morris}. Since the estimation of
the SgrA* three parameters is of significant importance for the
understanding of more general black holes, more restrictions are obtained on
their values using the available observational data \cite{Eisenhauer}\cite%
{Doeleman}\cite{Eckart}.

In a recent work \cite{Aguilar}, the Kerr black hole's parameters were
investigated by applying a relativistic stationary axisymmetric formalism 
\cite{Aguilar2} to the study of black hole rotation curves. The expressions
for the mass $M$, the rotation parameter a ($a=J/M$) and the distance
between the black hole center and a distant observer were obtained in terms
of the red/blue shifts of photons traveling along null geodesics and emitted
by massive objects orbiting the black hole in a stable circular equatorial
motion. These expressions of red/blue shifts allow to statistically estimate
the Kerr black hole parameter. However, this study do not take into account
the charge of the black hole.

Therefore, the purpose of this work is to investigate the parameters of a
charged rotating black hole, following the same formalism used in \cite%
{Aguilar}. We will use the Kerr-Newman metric to model this black hole, and
obtain the expressions of the red/blue shifts of photons emitted by massive
objects (a star, gas or dust) orbiting the black hole in the equatorial
plane, and traveling along null geodesics towards a distant observer.

This paper is organized as follows: Sec. 2 is devoted to the Kerr-Newman
metric, its corresponding conserved quantities and the equations of motion
for a neutral massive objects (stars, gas or dust) orbiting the Kerr-Newman
black hole. Then, we consider the special case of a circular motion in the
equatorial plane of the black hole. In Sec. 3 we derive the equations of
motions of photons traveling along null geodesics in the Kerr-Newman metric.
In Sec. 4 the expressions of red/blue shifts for photons emitted by neutral
massive objects in circular equatorial motion around the black hole are
obtained in terms of the emitter and detectors 4-velocities, angular
velocities and the photons impact parameter. The expressions of the red/blue
shifts in terms of the black hole parameters $M$, $a$, $Q$, the emitter and
detector radial coordinates $r_{e}$ and $r_{d}$ respectively are obtained in
Sec. 5. Having those expressions at hand, we can explore the behavior of
red/blue shifts in some special interesting cases like $Q=0$, $a=0$, and
when the detector is located very far away from Earth. Moreover, we get the
black hole angular parameter $a$ in terms of its mass $M$, charge $Q$ and
emitter radius $r_{e}$. These results are discussed further in Sec. 6.

\section{The Kerr-Newman metric}

The line element of the Kerr-Newman metric in the Boyer-Lindquist
coordinates is 
\begin{equation*}
ds^{2}=g_{tt}dt^{2}+2g_{t\phi }dtd\phi +g_{\phi \phi }d\phi
^{2}+g_{rr}dr^{2}+g_{\theta \theta }d\theta ^{2}\text{ \ },
\end{equation*}%
where

\begin{eqnarray}
g_{tt} &=&-\left( 1-\frac{2Mr-Q^{2}}{\Sigma }\right) \text{,} \\
g_{\phi t} &=&g_{t\phi }=-a\sin ^{2}\theta \left( \frac{2Mr-Q^{2}}{\Sigma }%
\right) \text{, \ \ \ }g_{rr}=\frac{\Sigma }{\Delta }\text{, \ \ }g_{\theta
\theta }=\Sigma \text{ ,} \\
\text{ \ \ }g_{\phi \phi } &=&\sin ^{2}\theta \left( r^{2}+a^{2}-\frac{%
a^{2}\left( 2Mr-Q^{2}\right) \sin ^{2}\theta }{\Sigma }\right)  \label{g}
\end{eqnarray}%
with

\begin{equation}
\Delta \equiv r^{2}+a^{2}-2Mr+Q^{2}\ ,\ \ \ \ \ \ \ \ \ \ \Sigma \equiv
r^{2}+a^{2}\cos ^{2}\theta \ ,\ \ \ \ \ \ \ \ a\equiv \frac{J}{M}\ .
\label{delt_sigm}
\end{equation}%
The Kerr-Newman black hole horizon condition is%
\begin{equation}
M^{2}\geq a^{2}+Q^{2}\ .
\end{equation}%
A\ very useful relation between the metric components is%
\begin{equation}
g_{t\phi }^{2}-g_{tt}g_{\phi \phi }=\sin ^{2}\theta \Delta
=r^{2}+a^{2}-2Mr+Q^{2}\text{ }
\end{equation}

The fact that Kerr-Newman metric is independent of the coordinates $t$ and $%
\phi $ implies the existence of the following Killing vectors%
\begin{equation}
\xi ^{\mu }=(1,0,0,0)\text{ \ },
\end{equation}%
\begin{equation}
\psi ^{\mu }=(0,0,0,1)\text{ \ }.
\end{equation}%
This results in the existence of two conserved quantities, the total energy $%
E$ and the angular momentum $L$. In addition, the Kerr metric possesses a
Killing tensor field \cite{Adamo}, implying the existence of another
conserved quantity, the Carter constant, denoted by $C$. Thus, for a neutral
test particle of mass $\mu $ moving in the Kerr-Newman black hole, the three
constants of motion are 
\begin{subequations}
\begin{eqnarray}
E &=&-g_{\mu \nu }\xi ^{\mu }P^{\nu }=-P_{t}=total\ energy,  \label{cons_t}
\\
L &=&g_{\mu \nu }\psi ^{\mu }P^{\nu }=P_{\phi }=angular\ momentum,
\label{cons_phi} \\
C &=&P_{\theta }^{2}+\cos ^{2}\theta \left[ a^{2}\left( \mu
^{2}-P_{t}^{2}\right) +\frac{P_{\phi }^{2}}{\sin ^{2}\theta }\right] 
\label{cons_theta} \\
&=&P_{\theta }^{2}+\cos ^{2}\theta \left[ a^{2}\left( \mu ^{2}-E^{2}\right) +%
\frac{L^{2}}{\sin ^{2}\theta }\right] \text{ \ \ .}
\end{eqnarray}%
Along side with the equation of normalization of the particle's momentum 
\end{subequations}
\begin{equation}
P^{\nu }P_{\nu }=P^{t}P_{t}+P^{r}P_{r}+P^{\theta }P_{\theta }+P^{\phi
}P_{\phi }=-\mu ^{2}\text{ \ \ },
\end{equation}%
the previous equations constitute the equations of motions of the particle.
After simplifications, the equations of motion of a neutral test particle in
the Kerr-Newman Black hole can be written as follows%
\begin{eqnarray}
\Sigma P^{t} &=&-a\left( a\sin ^{2}\theta E-L\right) +\left(
r^{2}+a^{2}\right) \frac{T}{\Delta }\text{ \ ,}  \label{Ea_motion_t} \\
\Sigma P^{r} &=&\left[ T^{2}-\Delta \left( \mu ^{2}r^{2}+\left( aE-L\right)
^{2}+C\right) \right] ^{\frac{1}{2}}\equiv V\left( r\right) ,
\label{Ea_motion_r} \\
\Sigma P^{\theta } &=&\left[ C-\cos ^{2}\theta \left( a^{2}\left( \mu
^{2}-E^{2}\right) +\frac{L^{2}}{\sin ^{2}\theta }\right) \right] ^{1/2}\text{
}\equiv \Theta (\theta ),  \label{Ea_motion_theta} \\
\Sigma P^{\phi } &=&-\left( aE-\frac{L}{\sin ^{2}\theta }\right) +a\frac{T}{%
\Delta }\equiv \Phi (r,\theta ),  \label{Ea_motion_phi}
\end{eqnarray}%
where $T=\left( r^{2}+a^{2}\right) E-aL$, while $V\left( r\right) $, $\Theta
(\theta )$ and $\Phi (r,\theta )$ are functions of coordinates.

\subsection{Circular equatorial motion of neutral massive test particles}

In order to obtain an explicit expression for the energy and angular
momentum of the emitter body (a star, gas or dust), we will examine the
neutral massive test particles in the equatorial plane ($\theta =\pi /2$).
If we require the particle to remain in that plane, which means $P^{\theta
}=\mu \left( d\theta /d\tau \right) =0$, where $\tau $ is the proper time,
the equation (\ref{Ea_motion_theta}) leads to $C=0$. Moreover, for circular
orbits in the equatorial plane at some radius $r$, the radial velocity $%
dr/d\tau =P^{r}/\mu $ must vanish instantaneously and at subsequent times.
Hence, equation (\ref{Ea_motion_r}) gives the following conditions 
\begin{equation}
V\left( r\right) =0\text{ \ \ \ and \ \ \ }\dot{V}\left( r\right) =0\text{ \
\ \ \ \ \ (for circular orbits),}  \label{circ_orb}
\end{equation}%
where the dot denotes derivative with respect to the radial coordinate.
Solving these equations for $E$ and $L$ yields the expression of the energy
and angular momentum for neutral massive test particle in the Kerr-Newman
spacetime \cite{Naresh}%
\begin{eqnarray}
E/\mu &=&\frac{r^{2}-2Mr+Q^{2}\pm a\left( Mr-Q^{2}\right) ^{\frac{1}{2}}}{r%
\left[ r^{2}-3Mr+2Q^{2}\pm 2a\left( Mr-Q^{2}\right) ^{\frac{1}{2}}\right]
^{1/2}}\text{ \ },  \label{energy} \\
L/\mu &=&\pm \frac{\left( Mr-Q^{2}\right) ^{\frac{1}{2}}\left[
r^{2}+a^{2}\mp 2a\left( Mr-Q^{2}\right) ^{\frac{1}{2}}\right] \mp aQ^{2}}{r%
\left[ r^{2}-3Mr+2Q^{2}\pm 2a\left( Mr-Q^{2}\right) ^{\frac{1}{2}}\right]
^{1/2}}\text{\ },  \label{ang_m}
\end{eqnarray}%
where the (+) sign correspond to a co-rotating test particle with respect to
the black hole rotation, while the (-) sign correspond to a counter rotating
one.

\section{Equations of motion for photons}

After discussing the emitter body, let's consider now the photons traveling
along null geodesics and emitted by these bodies. The equations of motion
for these photons in the Kerr-Newman metric can be obtained in a similar way
to those of massive particles. The only difference is the normalization
equation $k^{\nu }k_{\nu }=0$, where $k^{\nu }$is the momentum four-vector
of the photon. Hence, the constants of motion for photons are 
\begin{subequations}
\begin{eqnarray}
E_{\gamma } &=&-g_{\mu \nu }\xi ^{\mu }k^{\nu }=-k_{t}=total\ energy,
\label{cons_gam_t} \\
L_{\gamma } &=&g_{\mu \nu }\psi ^{\mu }k^{\nu }=k_{\phi }=angular\ momentum,
\label{cons_gam_phi} \\
C &=&k_{\theta }^{2}+\cos ^{2}\theta \left[ -a^{2}E_{\gamma }^{2}+\frac{%
L_{\gamma }^{2}}{\sin ^{2}\theta }\right] \text{ \ },  \label{cons_gam_theta}
\\
k^{\nu }k_{\nu } &=&k^{t}k_{t}+k^{r}k_{r}+k^{\theta }k_{\theta }+k^{\phi
}k_{\phi }=0\text{ \ }.  \label{k_norm}
\end{eqnarray}%
These equations can be written in the following simpler form 
\end{subequations}
\begin{subequations}
\begin{eqnarray}
\Sigma k^{t} &=&-a\left( aE_{\gamma }\sin ^{2}\theta -L_{\gamma }\right)
+\left( r^{2}+a^{2}\right) \frac{T_{\gamma }}{\Delta }\text{ \ }, \\
\Sigma k^{r} &=&\left[ T_{\gamma }^{2}-\Delta \left( \left( E_{\gamma
}a-L_{\gamma }\right) ^{2}+C\right) \right] ^{1/2}\text{ \ }, \\
\Sigma k^{\theta } &=&\left[ C-\cos ^{2}\theta \left( -a^{2}E_{\gamma }^{2}+%
\frac{L_{\gamma }^{2}}{\sin ^{2}\theta }\right) \right] ^{1/2}\text{ \ }, \\
\Sigma k^{\phi } &=&-\left( aE_{\gamma }-\frac{L_{\gamma }}{\sin ^{2}\theta }%
\right) +a\frac{T_{\gamma }}{\Delta }\text{ \ },
\end{eqnarray}%
where $T_{\gamma }=E_{\gamma }\left( r^{2}+a^{2}\right) -E_{\gamma }a$.

\section{Kinematic red/blue shifts of photons emitted by neutral massive
objects in equatorial circular motion}

\subsection{General expression of red/blue shifts}

Now we will find the general expression for red/blue shift of photons in the
Kerr-Newman spacetime. The frequency of a photon of momentum 4-vector $%
k^{\nu }$, measured by an observer with 4-velocity $U^{\nu }$ at point $P$
is given by 
\end{subequations}
\begin{equation}
\omega =-k_{\mu }U^{\mu }|_{P}\text{ \ }.
\end{equation}%
Thus, the frequencies of photons measured by an observer at the emission
point ($e$) and the detection point ($d$) respectively are%
\begin{eqnarray}
\omega _{e} &=&-k_{\mu }U^{\mu }|_{e}\text{ \ ,} \\
\omega _{d} &=&-k_{\mu }U^{\mu }|_{d}\text{ \ .}
\end{eqnarray}%
Hence, the expression of the frequency shift \emph{experienced} by photons
between the emission and detection points is%
\begin{eqnarray}
1+z &=&\frac{\omega _{e}}{\omega _{d}}=\frac{-k_{\mu }U^{\mu }|_{e}}{-k_{\mu
}U^{\mu }|_{d}}  \label{shift_gene} \\
&=&\frac{\left( E\gamma U^{t}-k_{r}U^{r}-k_{\theta }U^{\theta }-L\gamma
U^{\phi }\right) |_{e}}{\left( E\gamma U^{t}-k_{r}U^{r}-k_{\theta }U^{\theta
}-L\gamma U^{\phi }\right) |_{d}}  \notag \\
&=&\frac{\left( Ek^{t}-U_{r}k^{r}-U_{\theta }k^{\theta }-Lk^{\phi }\right)
|_{e}}{\left( Ek^{t}-U_{r}k^{r}-U_{\theta }k^{\theta }-Lk^{\phi }\right)
|_{d}}\text{ \ },  \notag
\end{eqnarray}%
where we have used equations (\ref{cons_gam_t},\ref{cons_gam_phi}) for the
constants $E\gamma $ and $L\gamma $ in the second line, and equations (\ref%
{cons_t},\ref{cons_phi}) for the constants $E$ and $L$ in the third
line.\bigskip\ This is the most general expression of frequency shift of
photons emitted and detected at tow different points ($e$) and ($d$), and
traveling along null geodesics. In the following, we will examine
specifically photons emitted by particles in circular orbits.

\subsection{Kinematic red/blue shifts of photons emitted by neutral massive
objects in equatorial circular motion}

From now on, we shall restrict ourselves to the frequency shift of photons
emitted by neutral massive particles in circular equatorial movements around
the black hole center with:

\begin{itemize}
\item Circular orbit: $U^{r}=0$ and $\dot{U}^{r}=0$

\item Equatorial plan: $\theta =\pi /2$, \ \ \ $U^{\theta }=0,$ \ $C=0$.
\end{itemize}

Therefore, the previously obtained expression of frequency shift, equation (%
\ref{shift_gene}), takes the form%
\begin{equation}
1+z=\frac{\omega _{e}}{\omega _{d}}=\frac{\left( E\gamma U^{t}-L\gamma
U^{\phi }\right) |_{e}}{\left( E\gamma U^{t}-L\gamma U^{\phi }\right) |_{d}}=%
\frac{U_{e}^{t}-b_{e}U_{e}^{\phi }}{U_{d}^{t}-b_{d}U_{d}^{\phi }}\ ,
\label{shift_rot_ph}
\end{equation}%
where we have introduced the impact parameter $b\equiv L\gamma /E\gamma $.
Since the quantities $L\gamma $ and $E\gamma $ are conserved along the
photons trajectory, which is a geodesic, the impact parameter $b$ is
conserved as well, so it has the same value in the points of emission and
detection respectively, i.e. $b_{e}=b_{d}=b$.

Now, let's consider the frequency shift $z_{c}$ due to the gravitational
field and an emitter having $b=0$. Substituting $b=0$ in equation (\ref%
{shift_rot_ph}), we obtain 
\begin{equation}
1+z_{c}=\frac{U_{e}^{t}}{U_{d}^{t}}\text{ \ .}
\end{equation}%
Thus, the kinematic frequency shift $z_{kin}$ is defined as the difference
between the total shift and the gravitational shift \cite{Aguilar}%
\begin{equation}
z_{kin}\equiv z-z_{c}\text{ \ \ ,}
\end{equation}%
where $z$ is defined by equation (\ref{shift_rot_ph}). Thus, the expression
of $z_{kin}$ is%
\begin{equation}
z_{kin}=\frac{U_{e}^{t}U_{d}^{\phi }b-U_{d}^{t}U_{e}^{\phi }b}{%
U_{d}^{t}\left( U_{d}^{t}-bU_{d}^{\phi }\right) }\text{ \ .}
\end{equation}%
The relevance of the kinematic shift is due to the fact that some
astronomers report their data in terms of $z_{kin}$ rather then the total
shift $z$.

As for the expression of the impact parameter $b$, we will consider the
photons emitted by objects orbiting either sides of the center of the source
and whose position vector $r$ with respect to the black hole center is
orthogonal to the detector's line of sight. Hence, these photons will
examine the maximum and minimum frequency shifts: a blue shift $z_{1}$ and
red shift $z_{2}$ emitted respectively to an approaching and receding
object, with respect to a far away positioned observer. Therefore, the
emitted photons has $k^{r}=k^{\theta }=0$. On one hand, substituting $%
k^{r}=k^{\theta }=0$ in equation (\ref{k_norm}) give%
\begin{equation}
b=\frac{L_{\gamma }}{E_{\gamma }}=\frac{k^{t}}{k^{\phi }}\text{ \ \ .}
\label{b_1}
\end{equation}%
On the other hand, using this expression in (\ref{k_norm}), then solving for 
$b$ yields 
\begin{equation}
b_{\pm }=-\frac{g_{t\phi }\pm \sqrt{g_{t\phi }^{2}-g_{tt}g_{\phi \phi }}}{%
g_{tt}}\text{ \ \ .}  \label{b2}
\end{equation}%
These two different values of $b$ give rise to the two different values of
the frequency shift $z_{1}$ and $z_{2}$, corresponding respectively to a
receding and an approaching object: 
\begin{subequations}
\label{z1_2}
\begin{eqnarray}
z_{1} &=&\frac{U_{e}^{t}U_{d}^{\phi }b_{-}-U_{d}^{t}U_{e}^{\phi }b_{-}}{%
U_{d}^{t}\left( U_{d}^{t}-b_{-}U_{d}^{\phi }\right) }\text{ \ \ ,} \\
z_{2} &=&\frac{U_{e}^{t}U_{d}^{\phi }b_{+}-U_{d}^{t}U_{e}^{\phi }b_{+}}{%
U_{d}^{t}\left( U_{d}^{t}-b_{+}U_{d}^{\phi }\right) }\text{ \ \ ,}
\end{eqnarray}%
For a detector located far away from the photons source, the angular
velocity is defined by 
\end{subequations}
\begin{equation}
\Omega _{d}\equiv \frac{d\phi }{dt}=\frac{d\phi /d\tau }{dt/d\tau }=\frac{%
U_{d}^{\phi }}{U_{d}^{t}}\text{ \ ,}  \label{ang_v}
\end{equation}%
where $\tau $ is the detector's proper time. Using this expression to
substitute $U_{d}^{\phi }$ in equations (\ref{z1_2}), we get \ 
\begin{eqnarray}
z_{1} &=&\frac{\left( U_{e}^{t}\Omega _{d}-U_{e}^{\phi }\right) b_{-}}{%
U_{d}^{t}\left( 1-\Omega _{d}b_{-}\right) }\text{ \ \ ,}  \label{z1} \\
z_{2} &=&\frac{\left( U_{e}^{t}\Omega _{d}-U_{e}^{\phi }\right) b_{+}}{%
U_{d}^{t}\left( 1-\Omega _{d}b_{+}\right) }\text{ \ \ .}  \label{z2}
\end{eqnarray}

\section{Expressions of red/blue shifts in terms of the Boyer-Lindquist
coordinates}

Now, we are going to find the expressions of the above frequency shift in
terms of $r_{d}$ and $r_{e}$, the radius of the emitter's and detector's
orbits, respectively.\ For this end, we need to write first the expressions
of $U_{d}^{t}$, $U_{d}^{\phi }$, $\Omega _{d}$ \ and $b_{\pm }$ in terms of
the radial coordinate.

Consider a massive neutral source of light of 4-velocity $U^{\mu }=U^{\mu
}/\tau $, orbiting the center of the black hole in a circular equatorial
plane ($\theta =\pi /2$). If the trajectory is required to be circular ($%
U^{r}=0$) and remain in the equatorial plane ($U^{\theta }=0$), the $t$
-component of the 4-velocity is%
\begin{equation}
U^{t}\left( r,\pi /2\right) =\frac{\left[ r^{4}+a^{2}r^{2}+a^{2}\left(
2Mr-Q^{2}\right) \right] }{r^{2}\left( r^{2}+a^{2}-2Mr+Q^{2}\right) }\left(
E/\mu \right) -\frac{a\left( 2Mr-Q^{2}\right) }{r^{2}\left(
r^{2}+a^{2}-2Mr+Q^{2}\right) }\left( L/\mu \right) \text{ \ ,}
\end{equation}%
whereas the $\phi $-component is%
\begin{equation}
U^{\phi }\left( r,\pi /2\right) =\frac{a\left( 2Mr-Q^{2}\right) }{%
r^{2}\left( r^{2}+a^{2}-2Mr+Q^{2}\right) }\left( E/\mu \right) +\frac{%
r^{2}-2Mr+Q^{2}}{r^{2}\left( r^{2}+a^{2}-2Mr+Q^{2}\right) }\left( L/\mu
\right) \text{ \ ,}
\end{equation}%
where equations (\ref{cons_t}) and (\ref{cons_phi}) were used. Now,
inserting (\ref{energy}) and (\ref{ang_m}) in the previous couple of
equations gives%
\begin{equation}
U^{t}\left( r,\pi /2\right) =\frac{r^{2}\pm a\left( Mr-Q^{2}\right) ^{\frac{1%
}{2}}}{r\left[ r^{2}-3Mr+2Q^{2}\pm 2a\left( Mr-Q^{2}\right) ^{\frac{1}{2}}%
\right] ^{1/2}}\text{ \ ,}  \label{U_t}
\end{equation}%
\begin{equation}
U^{\phi }\left( r,\pi /2\right) =\frac{\pm \left( Mr-Q^{2}\right) ^{\frac{1}{%
2}}}{r\left[ r^{2}-3Mr+2Q^{2}\pm 2a\left( Mr-Q^{2}\right) ^{\frac{1}{2}}%
\right] ^{1/2}}\ \text{\ .}  \label{U_phi}
\end{equation}

With the formulas of $U^{t}$ and $U^{\phi }$ at hand, it's straightforward
to write the angular velocity of the source of light which is moving in a
circular orbit in the equatorial plane, with respect to an observer. Hence,
equation (\ref{ang_v}), in terms of the Boyer-Lindquist coordinates%
\begin{equation}
\Omega =\frac{\pm \left( Mr-Q^{2}\right) ^{\frac{1}{2}}}{r^{2}\pm a\left(
Mr-Q^{2}\right) ^{\frac{1}{2}}}  \label{ang_v2}
\end{equation}%
where the (+) and (-) signs respectively correspond to co-rotating and
counter-rotating objects with respect to the black hole angular momentum.

In addition to that, the expression of the impact parameter $b_{\pm }$,
equation (\ref{b2}), in terms of the Boyer-Lindquist coordinates (with $%
\theta =\pi /2$) is 
\begin{equation}
b_{\pm }=\frac{-a\left( 2Mr-Q^{2}\right) \pm r^{2}\sqrt{r^{2}+a^{2}-2Mr+Q^{2}%
}}{r^{2}-2Mr+Q^{2}}\text{ \ \ .}  \label{b3}
\end{equation}%
\bigskip

Having done all this, we can now write the frequency shift in terms of the
Boyer-Lindquist coordinate $r$. Inserting equations (\ref{U_t}) and (\ref%
{U_phi}) in (\ref{z1}) yields%
\begin{equation}
z_{1}=\frac{r_{d}\left[ r_{d}^{2}-3Mr_{d}+2Q^{2}\pm 2a\left(
Mr_{d}-Q^{2}\right) ^{\frac{1}{2}}\right] ^{1/2}}{r_{e}\left[
r_{e}^{2}-3Mr_{e}+2Q^{2}\pm 2a\left( Mr_{e}-Q^{2}\right) ^{\frac{1}{2}}%
\right] ^{1/2}}\frac{\left( \left[ r_{e}^{2}\pm a\left( Mr_{e}-Q^{2}\right)
^{\frac{1}{2}}\right] \Omega _{d}-\pm \left( Mr_{e}-Q^{2}\right) ^{\frac{1}{2%
}}\right) b_{-}}{\left[ r_{d}^{2}\pm a\left( Mr_{d}-Q^{2}\right) ^{\frac{1}{2%
}}\right] \left( 1-\Omega _{d}b_{-}\right) }\text{ \ ,}
\end{equation}%
where $r_{d}$ and $r_{e}$ represent the radius of the emitter and detector's
orbits, respectively. Using equation (\ref{ang_v2}), we can get the
following simpler expression 
\begin{equation}
z_{red}=\frac{r_{d}\left[ r_{d}^{2}-3Mr_{d}+2Q^{2}\pm 2a\left(
Mr_{d}-Q^{2}\right) ^{\frac{1}{2}}\right] ^{1/2}}{r_{e}\left[
r_{e}^{2}-3Mr_{e}+2Q^{2}\pm 2a\left( Mr_{e}-Q^{2}\right) ^{\frac{1}{2}}%
\right] ^{1/2}}\times \frac{\Omega _{d}\left( \Omega _{d}-\Omega _{e}\right)
b_{-}}{\Omega _{e}\left( 1-\Omega _{d}b_{-}\right) }\times \left[ \frac{%
Mr_{e}-Q^{2}}{Mr_{d}-Q^{2}}\right] ^{1/2}\text{ \ .}  \label{z1_b}
\end{equation}%
Following the same steps with $z_{2}$,\ gives%
\begin{equation}
z_{blue}=\frac{r_{d}\left[ r_{d}^{2}-3Mr_{d}+2Q^{2}\pm 2a\left(
Mr_{d}-Q^{2}\right) ^{\frac{1}{2}}\right] ^{1/2}}{r_{e}\left[
r_{e}^{2}-3Mr_{e}+2Q^{2}\pm 2a\left( Mr_{e}-Q^{2}\right) ^{\frac{1}{2}}%
\right] ^{1/2}}\times \frac{\Omega _{d}\left( \Omega _{d}-\Omega _{e}\right)
b_{+}}{\Omega _{e}\left( 1-\Omega _{d}b_{+}\right) }\times \left[ \frac{%
Mr_{e}-Q^{2}}{Mr_{d}-Q^{2}}\right] ^{1/2}\ \ .\bigskip   \label{z2_b}
\end{equation}%
Interestingly, notice that there is no frequency shift if:

\begin{itemize}
\item the emitter and detector angular velocities are the same $\Omega
_{d}=\Omega _{e}$.

\item the charge of the Kerr-Newman black hole takes the value $%
Q^{2}=Mr_{e}. $
\end{itemize}

we obtain the expression of the red shift $z_{red}$ in terms of the
Kerr-Newman black hole parameters $M$, $a$, $Q$ and the detector radius $%
r_{d}$.

We can further substitute the expressions of $\Omega $ and $b$ in $z_{red}$
and $z_{blue}$. Using equations (\ref{ang_v2}) and (\ref{b3}), the second
factor of (\ref{z1_b}) takes the form%
\begin{eqnarray}
\frac{\Omega _{d}\left( \Omega _{d}-\Omega _{e}\right) b_{-}}{\Omega
_{e}\left( 1-\Omega _{d}b_{+}\right) }\text{ } &\text{=}&\pm \left[ \frac{%
Mr_{d}-Q^{2}}{Mr_{e}-Q^{2}}\right] ^{1/2}\times \frac{r_{d}^{2}\left(
Mr_{e}-Q^{2}\right) ^{\frac{1}{2}}-r_{e}^{2}\left( Mr_{d}-Q^{2}\right) ^{%
\frac{1}{2}}}{\left[ r_{d}^{2}\pm a\left( Mr_{d}-Q^{2}\right) ^{\frac{1}{2}}%
\right] }  \notag \\
&&\times \frac{\left[ a\left( 2Mr_{e}-Q^{2}\right) +r_{e}^{2}\sqrt{\Delta
(r_{e})}\right] }{r_{d}^{2}\left( r_{e}^{2}-2Mr_{e}+Q^{2}\right) \pm \left(
Mr_{d}-Q^{2}\right) ^{\frac{1}{2}}r_{e}^{2}\left[ a+\sqrt{\Delta (r_{e})}%
\right] }\text{ \ .}
\end{eqnarray}%
Inserting this relation in (\ref{z1_b}) yields an expression for the
kinematic frequency shift $z_{red}$ that depends only on the Kerr-Newman
black hole parameters $a$ , $M$, $Q,$ and the detector's and emitter's radii 
$r_{d}$ and $r_{e}$ respectively%
\begin{eqnarray}
z_{red} &=&\pm \frac{r_{d}\left[ r_{d}^{2}-3Mr_{d}+2Q^{2}\pm 2a\left(
Mr_{d}-Q^{2}\right) ^{\frac{1}{2}}\right] ^{1/2}}{r_{e}\left[
r_{e}^{2}-3Mr_{e}+2Q^{2}\pm 2a\left( Mr_{e}-Q^{2}\right) ^{\frac{1}{2}}%
\right] ^{1/2}}\times \frac{r_{d}^{2}\left( Mr_{e}-Q^{2}\right) ^{\frac{1}{2}%
}-r_{e}^{2}\left( Mr_{d}-Q^{2}\right) ^{\frac{1}{2}}}{\left[ r_{d}^{2}\pm
a\left( Mr_{d}-Q^{2}\right) ^{\frac{1}{2}}\right] }  \notag \\
&&\times \frac{\left[ a\left( 2Mr_{e}-Q^{2}\right) +r_{e}^{2}\sqrt{\Delta
(r_{e})}\right] }{r_{d}^{2}\left( r_{e}^{2}-2Mr_{e}+Q^{2}\right) \pm
r_{e}^{2}\left( Mr_{d}-Q^{2}\right) ^{\frac{1}{2}}\left[ a+\sqrt{\Delta
(r_{e})}\right] }\text{ \ \ ,}  \label{z1_c}
\end{eqnarray}%
where $\Delta (r)=r^{2}+a^{2}-2Mr+Q^{2}$. Following the same steps, we find
a similar expression for $z_{2}$%
\begin{eqnarray}
z_{blue} &=&\pm \frac{r_{d}\left[ r_{d}^{2}-3Mr_{d}+2Q^{2}\pm 2a\left(
Mr_{d}-Q^{2}\right) ^{\frac{1}{2}}\right] ^{1/2}}{r_{e}\left[
r_{e}^{2}-3Mr_{e}+2Q^{2}\pm 2a\left( Mr_{e}-Q^{2}\right) ^{\frac{1}{2}}%
\right] ^{1/2}}\times \frac{r_{d}^{2}\left( Mr_{e}-Q^{2}\right) ^{\frac{1}{2}%
}-r_{e}^{2}\left( Mr_{d}-Q^{2}\right) ^{\frac{1}{2}}}{\left[ r_{d}^{2}\pm
a\left( Mr_{d}-Q^{2}\right) ^{\frac{1}{2}}\right] }  \notag \\
&&\times \frac{\left[ a\left( 2Mr_{e}-Q^{2}\right) -r_{e}^{2}\sqrt{\Delta
(r_{e})}\right] }{r_{d}^{2}\left( r_{e}^{2}-2Mr_{e}+Q^{2}\right) \pm
r_{e}^{2}\left( Mr_{d}-Q^{2}\right) ^{\frac{1}{2}}\left[ a-\sqrt{\Delta
(r_{e})}\right] }  \label{z2_c}
\end{eqnarray}%
As expected, these expressions reduce to those obtained for the Kerr black
hole\ when we put $Q=0$ in (\ref{z1_c}) and (\ref{z2_c}):%
\begin{eqnarray}
z_{red} &=&\pm M^{^{\frac{1}{2}}}\frac{r_{d}^{3/4}\left[
r_{d}^{3/2}-3Mr_{d}^{1/2}\pm 2aM^{^{1/2}}\right] ^{1/2}}{r_{e}^{3/4}\left[
r_{e}^{3/2}-3Mr_{e}^{1/2}\pm 2aM^{^{1/2}}\right] ^{1/2}}\times \frac{%
r_{d}^{3/2}-r_{e}^{3/2}}{\left[ r_{d}^{3/2}\pm aM^{^{1/2}}\right] }  \notag
\\
&&\times \frac{\left[ a\left( 2M\right) +r_{e}\sqrt{r_{e}^{2}+a^{2}-2Mr_{e}}%
\right] }{r_{d}^{3/2}\left( r_{e}-2M\right) \pm ar_{e}M^{^{1/2}}\pm
r_{e}M^{^{1/2}}\sqrt{r_{e}^{2}+a^{2}-2Mr_{e}}}\text{ \ \ ,}
\end{eqnarray}%
\begin{eqnarray}
z_{blue} &=&\pm M^{^{\frac{1}{2}}}\frac{r_{d}^{3/4}\left[
r_{d}^{3/2}-3Mr_{d}^{1/2}\pm 2aM^{^{1/2}}\right] ^{1/2}}{r_{e}^{3/4}\left[
r_{e}^{3/2}-3Mr_{e}^{1/2}\pm 2aM^{^{1/2}}\right] ^{1/2}}\times \frac{%
r_{d}^{3/2}-r_{e}^{3/2}}{\left[ r_{d}^{3/2}\pm aM^{^{1/2}}\right] }  \notag
\\
&&\times \frac{\left[ a\left( 2M\right) -r_{e}\sqrt{r_{e}^{2}+a^{2}-2Mr_{e}}%
\right] }{r_{d}^{3/2}\left( r_{e}-2M\right) \pm ar_{e}M^{^{1/2}}\mp
r_{e}M^{^{1/2}}\sqrt{r_{e}^{2}+a^{2}-2Mr_{e}}}\text{ \ \ ,}
\end{eqnarray}%
which are exactly the expressions obtained in \cite{Aguilar}.

Furthermore, if we take the static case $a=J/M=0$ of the Kerr-Newman black,
we get the red/blue shift expressions for a charged nonrotating black hole%
\begin{eqnarray}
z_{red} &=&\pm \frac{r_{e}}{r_{d}}\left[ \frac{r_{d}^{2}-3Mr_{d}+2Q^{2}}{%
r_{e}^{2}-3Mr_{e}+2Q^{2}}\right] ^{1/2}\times \frac{r_{d}^{2}\left(
Mr_{e}-Q^{2}\right) ^{\frac{1}{2}}-r_{e}^{2}\left( Mr_{d}-Q^{2}\right) ^{%
\frac{1}{2}}}{r_{d}^{2}\left( r_{e}^{2}-2Mr_{e}+Q^{2}\right) ^{1/2}\pm
r_{e}^{2}\left( Mr_{d}-Q^{2}\right) ^{\frac{1}{2}}}\text{\ \ \ ,} \\
z_{red} &=&\mp \frac{r_{e}}{r_{d}}\left[ \frac{r_{d}^{2}-3Mr_{d}+2Q^{2}}{%
r_{e}^{2}-3Mr_{e}+2Q^{2}}\right] ^{1/2}\times \frac{r_{d}^{2}\left(
Mr_{e}-Q^{2}\right) ^{\frac{1}{2}}-r_{e}^{2}\left( Mr_{d}-Q^{2}\right) ^{%
\frac{1}{2}}}{r_{d}^{2}\left( r_{e}^{2}-2Mr_{e}+Q^{2}\right) ^{1/2}\mp
r_{e}^{2}\left( Mr_{d}-Q^{2}\right) ^{\frac{1}{2}}}\text{\ \ \ .}
\end{eqnarray}%
Therefore, these are expressions of frequency shifts for of photons
traveling along null geodesics, emitted by neutral massive objects orbiting
a Reissner-Nordstr\"{o}m black hole in circular equatorial motion at radius $%
r_{e}$, and detected at radius $r_{d}$.

\subsection{Equation for $r_{d}$ in terms of $a$, $Q$ and $b_{e}$}

The fact that $E_{\gamma }$ and $L_{\gamma }$ are constant along the photons
path imply that the impact parameter $b$ is constant too. In particular, $b$
has the same value regardless of being measured at the emitter or detector
position: $b_{e}=b_{d}$. Using (\ref{b3}), we get%
\begin{eqnarray}
&&r_{d}^{6}-2Mr_{d}^{5}+\left( -b_{e}+a^{2}+Q^{2}\right)
r_{d}^{4}+4b_{e}\left( b_{e}-a\right) r_{d}^{3}+2\left[ 2M^{2}\left(
b_{e}-a\right) ^{2}-b_{e}Q^{2}\left( b_{e}-a\right) \right] r_{d}^{2} \\
&&+4MQ^{2}\left( b_{e}-a\right) ^{2}r_{d}-Q^{4}\left( b_{e}-a\right) ^{2}=0%
\text{ \ .}  \notag
\end{eqnarray}%
Solving this equation will determine the value of $r_{d}$ in terms of $a$, $M
$, $Q$ and $b_{e}$.

\subsection{Expression of $z_{1}$ and $z_{2}$ for a far away observer}

An interesting case is when the detector is located far away from the source
so that $r_{d}\gg M\geq \sqrt{a^{2}+Q^{2}}$ (the Kerr-Newman black hole
horizon condition) and $r_{d}\gg r_{e}$, expressions of red/blue shifts take
the form%
\begin{equation}
z_{red}=\pm \frac{\left( Mr_{e}-Q^{2}\right) ^{\frac{1}{2}}}{r_{e}\left[
r_{e}^{2}-3Mr_{e}+2Q^{2}\pm 2a\left( Mr_{e}-Q^{2}\right) ^{\frac{1}{2}}%
\right] ^{1/2}}\times \frac{a\left( 2Mr_{e}-Q^{2}\right) +r_{e}^{2}\sqrt{%
\Delta (r_{e})}}{\left( r_{e}^{2}-2Mr_{e}+Q^{2}\right) }  \label{z1_d}
\end{equation}%
\bigskip 
\begin{equation}
z_{blue}=\pm \frac{\left( Mr_{e}-Q^{2}\right) ^{\frac{1}{2}}}{r_{e}\left[
r_{e}^{2}-3Mr_{e}+2Q^{2}\pm 2a\left( Mr_{e}-Q^{2}\right) ^{\frac{1}{2}}%
\right] ^{1/2}}\times \frac{a\left( 2Mr_{e}-Q^{2}\right) -r_{e}^{2}\sqrt{%
\Delta (r_{e})}}{\left( r_{e}^{2}-2Mr_{e}+Q^{2}\right) }  \label{z2_d}
\end{equation}

Since this couple of equations link the red/blue shifts to the Kerr-Newman
black hole parameters $M$, $a$ and $Q$, it will be very helpful if we can
directly express these parameters in terms of the frequency shifts (\ref%
{z1_d}) and (\ref{z2_d}). I we define $\alpha $ and $\beta $ as 
\begin{eqnarray}
\alpha  &\equiv &\left( z_{red}+z_{blue}\right) ^{2}\text{ \ ,}
\label{alpha} \\
\beta  &\equiv &\left( z_{red}-z_{blue}\right) ^{2}\text{ \ ,}  \label{beta}
\end{eqnarray}%
we get the following expression for the rotation parameter%
\begin{equation}
a^{2}=\frac{\left( r_{e}^{6}-2Mr_{e}^{5}+Q^{2}r_{e}^{4}\right) \alpha }{%
-\alpha r_{e}^{4}+\beta \left( 4M^{2}r_{e}^{2}+Q^{4}-4MQ^{2}r_{e}\right) }%
\text{ \ .}
\end{equation}

Notice that if $z_{red}=z_{blue}$, we get 
\begin{eqnarray*}
a^{2} &=&-\left( r_{e}^{2}-2Mr_{e}+Q^{2}\right)  \\
&\rightarrow &r_{e}^{2}-+a^{2}+2Mr_{e}+Q^{2}=\Delta =0\text{ \ .}
\end{eqnarray*}%
which is the equation of Kerr-Newman horizons, namely $g^{rr}=\frac{\Delta }{%
\Sigma }=0$. This means that the condition $z_{red}=z_{blue}$ can only
happen when the radius of the emitter $r_{e}$ is near the black hole
horizons.

\section{Conclusion}

In this work, we determined the mass,\ rotation, and charge parameters of
the Kerr-Newman black hole in terms\ of red/blue shifts of photons emitted
by geodesic particles. We obtained an explicit expression for the energy and
angular momentum of the emitted body, and derived the equations of motion
for photons in the Kerr-Newman metric. We established the form of red/blue
shifts of photons emitted by neutral massive objects orbiting a Kerr-Newman
black hole incircular equatorial motion. Using Boyer-Lindquist coordinates,
we expressed the red/blue shifts in terms of the Kerr-Newman metric
parameters. In the cases where the angular velocity of the detector vanishes
i.e. $\Omega _{d}=0$ or the charge of the Kerr-Newman black hole takes the
value $Q^{2}=Mr_{e}$, the mentioned red/blue shifts vanish. By setting $Q=0$%
, we recover the results given in \cite{Aguilar}. When the parameter $a=0$,
the study of red/blue shifts of emitted photons \ reduces to the
Reissner-Nordstr\"{o}m black hole. In the case where the detector is located
far away from the source so that $r_{d}\gg M\geq \sqrt{a^{2}+Q^{2}}$, if $%
z_{red}=z_{blue}$, the radius of the emitter $r_{e}$ is approximately equal
to the back hole horizons. Experimentally, these results allow us to make an
estimation of the red/blue shifts of emitted photons. This estimation, along
with the results presented in this paper, may lead to many interesting
applications in astrophysical phenomena related to black holes.

\end{document}